\documentclass[a4paper,fleqn,usenatbib]{mnras}
\usepackage[T1]{fontenc}
\usepackage{ae,aecompl}
\usepackage{graphicx}	% Including figure files
\usepackage{amsmath}	% Advanced maths commands
\usepackage{amssymb}	% Extra maths symbols
\title[Electromagnetic power from a rotating black hole]
{Outgoing electromagnetic power induced from pair plasma
falling into a rotating black hole}
\author[Y. Kojima]{Yasufumi Kojima
\thanks{%
E-mail:ykojima-phys@hiroshima-u.ac.jp}\\ 
Department of Physics, Hiroshima University, Higashi-Hiroshima, 
739-8526, Japan}
\date{\today}

\pubyear{2015}

\begin{document}
\label{firstpage}
\pagerange{\pageref{firstpage}--\pageref{lastpage}}
\maketitle

%
%%%%%%%%%%%%%%%%%%%%%%%%%%%%%%%%%%%%%%%%%%%%%%%%%%
% Abstract of the paper
\begin{abstract}
We examine energy conversion from accreting pair plasma to outgoing Poynting 
flux by black hole rotation. 
Our approach is based on a two-fluid model consisting of collisionless 
pair plasma. The electric potential is not constant along
magnetic field lines, unlike an ideal magnetohydrodynamics approximation. 
We show how and where longitudinal electric fields and toroidal 
magnetic fields are generated by the rotation, whereas
they vanish everywhere for radial flow in a
split monopole magnetic field in a Schwarzschild black hole.
Outgoing electromagnetic power in a steady state is calculated 
by applying the WKB method to the perturbation equations 
for a small spin parameter.
In our model, the luminosity has a peak in the vicinity of the
black hole, but is damped toward the event horizon and infinity.
The power at the peak is of the same order as that in the Blandford--Znajek process,
although the physical mechanism is different.
\end{abstract}
%%%%%%%%%%%%%%%%%%%%%%%%%%%%%%%%%%%%%%%%%%%%%%%%%%

%(1)%%%%%%%%%%%%%%%%%%%%%%%%%%%%%%%%%%%%%%%%%%%
\section{Introduction}
%%%%%%%%%%%%%%%%%%%%%%%%%%%%%%%%%%%%%%%%%%%%%%%
The Blandford--Znajek (BZ) process is widely discussed as a
promising mechanism for the powerful central engines in active galactic nuclei,
microquasars, and gamma ray bursts. In their seminal paper, 
\citet{1977MNRAS.179..433B}
showed outgoing energy flux from the event horizon 
on the assumption of a steady force-free magnetosphere around a
slowly rotating Kerr black hole. 
Owing to the simplified situation, the electromagnetic extraction 
of the rotational energy could be analytically demonstrated.
This remarkable process has been studied, partly because
the underlying assumptions are doubted and partly because realistic astrophysical relevance is very important.

Global steady-state force-free magnetospheres are 
constructed by numerically solving the relativistic Grad--Shafranov equation, in which there
are singular surfaces that make careful treatment necessary 
\citep[e.g.,][]{2004ApJ...603..652U,2005ApJ...620..889U,2013ApJ...765..113C,
2014ApJ...788..186N}.
Higher-order corrections to the original split monopole solution
are calculated with respect to the black hole spin
\citep[e.g.,][]{2008PhRvD..78b4004T,2015PhRvD..91f4067P}.
The Grad--Shafranov equation for magnetohydrodynamics (MHD) equilibrium is
much more complicated and difficult to solve \citep[e.g.,][]{1990ApJ...363..206T,
1991PhRvD..44.2295N,1993PhyU...36..529B}.
A time-dependent approach may thus be preferable,
and general relativistic MHD
simulations provide very interesting models
\citep[e.g.,][]{2002Sci...295.1688K,2003ApJ...584..937V,
2004MNRAS.350.1431K,2005MNRAS.359..801K,
2006MNRAS.368.1561M,2009MNRAS.397.1153K,
2012MNRAS.423.3083M}.
Recently, very complicated but more realistic configurations
of the magnetic fields have been numerically treated.
Of particular interest is the dynamical process in gamma ray bursts.
The accretion of matter and jet emission in the vicinity 
of a central black hole can be explored simultaneously.
Sometimes, transient features are also exhibited in the numerical 
simulations. Outgoing flow from a system of a black hole coupled with surrounding 
magnetic fields has been tested, but
it is rather difficult to determine the most important elements
from numerical results. 

The mechanism for converting rotational energy to outgoing 
electromagnetic flux resembles that of a pulsar. However, 
the origin of electromotive force in a unipolar induction model 
is not established in the black hole magnetosphere.
Possibilities include 
the event horizon (\cite{1986bhmp.book.....T} and 
the critique by \cite{1990ApJ...350..518P}),
ergosphere \citep{2004MNRAS.350..427K,2014MNRAS.442.2855T},
and the pair creation surface
\citep{2000NCimB.115..795B,2012PASJ...64...50O,
2015PASJ..tmp..217O}
(see also, \citep{2008ASSL..355.....P,2010PhyU...53.1199B}
and the references therein).
When the ideal MHD condition holds in the entire magnetosphere,
the angular velocity~$\Omega_{\rm F} $ of the magnetic field is 
a function of magnetic flux, and this
characterizes the electric potential difference between 
magnetic field lines. 
This value is crucial in the BZ mechanism, which works in only a 
certain range of~$\Omega_{\rm F} $.
There remain the problems of where and how it is specified.
\citet{2014MNRAS.442.2855T}
argued that a breakdown of the ideal MHD condition in the 
ergosphere is essential for giving rise to electromotive force.
However, their argument is qualitative by an almost analytic 
treatment. It is impossible to discuss the origin of electromotive 
force in the framework of the ideal MHD, and it is 
necessary to study it beyond the approximation level.

In this paper, we do not assume ideal MHD conditions; we 
instead consider a two-component plasma consisting of positively and 
negatively charged particles, whose flows are governed by electromagnetic fields and gravity.
Maxwell's equations are solved with source terms of electric charge 
and current derived by the plasma motions, thus obtaining 
self-consistent solutions.
Our approach allows us to explore the origin of the electromotive force and 
outgoing electromagnetic flux, if it exists. 
There are few works applying a two-fluid model to astrophysical situations,
particularly to the formalism in 
black hole spacetime\citep{1998MNRAS.294..673K}, and stationary 
pulsar models\citep{2009MNRAS.398..271K,2015MNRAS.446.2243P}.
There are large numbers involved in the model to connect microscopic to macroscopic sizes, and this fact is an obstacle to numerical calculation.

  This paper is organized as follows.
Electromagnetic fields in a Kerr spacetime are discussed in Section~2
using~$3+1$ formalism. We provide a brief review,
because the equations in many papers assume
the ideal MHD condition and those without the condition are needed here.
We also discuss plasma flows
interacting with electromagnetic fields in curved spacetime.
The equations of stream functions are derived.
In Section~3, we present a model to investigate how
black hole spin modifies plasma flows and electromagnetic fields,
resulting in outgoing energy flux.
In a Schwarzschild spacetime, both the flow and magnetic field are radial.
That is, the magnetic field has a split-monopole configuration and the 
electric field vanishes.
To consider slow rotation of the back hole, we adopt the 
perturbation technique and explicitly obtain the results.
We present our conclusions in Section~4.

Note that in the following we assume axial symmetry and stationarity
in electromagnetic fields and plasma flows.
We use units of $c=G=1$.

%(2)%%%%%%%%%%%%%%%%%%%%%%%%%%%%%%%%%%%%%%%%%%%
\section{Basic equations}
%%%%%%%%%%%%%%%%%%%%%%%%%%%%%%%%%%%%%%%%%%%%%%%
%%%%%%%%%%%%%%%%%%%%%%%%%%
\subsection{Electromagnetic fields}
%%%%%%%%%%%%%%%%%%%%%%%%%%
In this section, we briefly summarize the Maxwell equations in a black hole 
spacetime. The Kerr metric in the Boyer-Lindquist coordinate is given by
\begin{equation}
ds^2=-\alpha^2dt^2+\frac{\rho^2}{\Delta}dr^2
       +\rho^2d\theta^2 +\varpi^2(d\phi-\omega dt)^2,
\end{equation}
where
\begin{eqnarray}
&&
\alpha^2=\frac{\rho^2\Delta}{\Sigma^2},
~ \rho^2=r^2+a^2\cos^2\theta,
~  \Delta=r^2+a^2-2Mr,
\nonumber
\\
&&  \varpi^2=\frac{\Sigma^2}{\rho^2}\sin^2\theta ,
~ \Sigma^2=(r^2+a^2)^2-a^2\Delta\sin^2\theta ,
\nonumber
\\
&&
 \omega=\frac{2Mar}{\Sigma^2}.
\end{eqnarray}

We consider stationary and axially symmetric fields, and the electromagnetic
vectors $\vec{E}$ and $\vec{B}$ refer to quantities measured by a locally 
non-rotating zero angular momentum observer (ZAMO).
The four-velocity $u^\mu$ of a ZAMO in Boyer--Lindquist coordinates is given by
%%%
\begin{equation}
u^\mu =\frac{dx^\mu}{d\tau} =
\left[ \frac{1}{\alpha}, -\frac{1}{\alpha}{\vec \beta } \right],
%~~{\vec \beta } =
[\beta^{\hat r}, \beta^{\hat \theta}, \beta^{\hat \phi}]
=[0,0, -\omega \varpi].
\end{equation}
In this paper, the notation ${\hat i}$ denotes the component of a vector
in an orthogonal basis.  Using vector analysis in a 3-dimensional 
curved space,
the time-independent Maxwell equations are expressed as
\citep[e.g.,][]{1986bhmp.book.....T}
\begin{equation}
\vec{\nabla}\cdot\vec{E}=4\pi\rho_{e},
\label{eqnMxw1}
\end{equation}
\begin{equation}
\vec{\nabla}\cdot\vec{B}=0,
\label{eqnMxw2}
\end{equation}
\begin{equation}
\vec{\nabla}\times(\alpha{\vec{E}})
=(\vec{\beta}\cdot\vec{\nabla}){\vec{B}}-(\vec{B}\cdot\vec{\nabla})
{\vec{\beta}},
\label{eqnMxw3}
\end{equation}
\begin{equation}
\vec{\nabla}\times(\alpha{\vec{B}})=
4\pi\alpha\vec{j}-(\vec{\beta}\cdot\vec{\nabla}){\vec{E}}
+(\vec{E}\cdot\vec{\nabla}){\vec{\beta}}.
\label{eqnMxw4}
\end{equation}
  It is convenient to introduce three scalar functions
$G(r,\theta)$, $S(r,\theta)$, and $\Phi(r,\theta)$ to express 
these fields. These respectively represent the magnetic flux,
poloidal current flow, and electric potential\footnote{
In the literature, symbols such as 
$\Phi $, $\Psi$, $A$, or $f$ are used instead of $G$, 
and the electric potential is related by the ideal MHD condition.
We do not assume this condition, so 
a new set of symbols is used here.
}.
The following forms satisfy eqs.~(\ref{eqnMxw2}) and (\ref{eqnMxw3}):
\begin{equation}
\vec{B}=\frac{\vec\nabla{G}\times\vec{e}_{\hat{\phi}}}{\varpi}
+\frac{S}{\alpha\varpi}\vec{e}_{\hat{\phi}},
\label{eqnDefBB}
\end{equation}
\begin{equation}
\vec{E}=-\frac{1}{\alpha}\vec{\nabla}\Phi
-\frac{\vec\beta}{\alpha}\times\vec{B}
=-\frac{1}{\alpha}
( \vec{\nabla}\Phi -\omega \vec{\nabla}G ).
\label{eqnDefEE}
\end{equation}
Their components are explicitly written as
\begin{equation}
[B_{\hat{r}},B_{\hat{\theta}}, B_{\hat{\phi}}]
=\left[\frac{1}{\varpi\rho}G,_\theta,
 -\frac{\Delta^{1/2}}{\varpi\rho}G,_r,
 \frac{S}{\alpha\varpi}\right],
\label{Bcomp}
\end{equation}
%%%%
\begin{equation}
[E_{\hat{r}},E_{\hat{\theta}}, E_{\hat{\phi}}]
=\left [
  -\frac{\Delta^{1/2}}{\alpha\rho}(\Phi,_r -\omega G,_r),
 -\frac{1}{\alpha\rho}(\Phi,_\theta - \omega G,_\theta),
 0 \right]
\label{Ecomp}
\end{equation}

Since $E^2-B^2 =((\omega\varpi/\alpha)^2 -1)|\nabla{G}/\varpi|^2$,
it is easily found that the electric fields dominate inside
the ergoregion if the toroidal magnetic fields and
longitudinal electric fields vanish ($S=\Phi=0$).
Equivalently, the magnitude of
`${\vec E} \times {\vec B}$' drift velocity $v _{d}$ 
exceeds the speed of light inside the ergoregion,
$ |v _{d}| = |{\vec E} \times {\vec B}|/B^2 $
$= |\omega\varpi/\alpha| >1 $. 
This indicates a breakdown of the MHD condition
\citep{2014MNRAS.442.2855T}.
As a result, longitudinal electric fields
arise to redistribute charge density.

The poloidal component $(r, \theta)$ of eq.~(\ref{eqnMxw4}) is given by
\begin{equation}
4\pi\alpha {\vec j}_{p}
=\frac{\vec\nabla{S}\times\vec{e}_{\hat{\phi}}}{\varpi}.
\label{enqjcpol}
\end{equation}
The poloidal current flows along a constant line $S$.
The toroidal component $(\phi) $ of eq.~(\ref{eqnMxw4}) is
\begin{equation}
\mathcal{D}G +
  \frac{\varpi^2 }{\alpha\rho^2} \left[
\omega_{,r}\Delta(\Phi_{,r}-\omega G_{,r})
+ \omega_{,\theta}( \Phi_{,\theta}-\omega G_{,\theta})\right] 
= 
-4\pi\alpha \varpi j_{\hat \phi} ,
\label{eqnBioG}
\end{equation}
where $\mathcal{D}$ is a differential operator given by
\begin{eqnarray}
&&
{\mathcal{D}}G = \varpi ^2 {\vec \nabla} \cdot \left(
\frac{\alpha}{\varpi ^2} {\vec \nabla} G \right)
\nonumber
\\
&&~~
=
\frac{\Delta^{1/2}\varpi }{\rho^2}\left[ 
\left( \frac{\alpha\Delta^{1/2}}{\varpi}
G_{,r} \right)_{,r}
+\left(\frac{\alpha}{\varpi\Delta^{1/2}} G_{,\theta}
\right)_{,\theta} \right] .
\end{eqnarray}
Finally, Gauss' law (\ref{eqnMxw1}) is explicitly written as
\begin{eqnarray}
&&
\frac{\Delta^{1/2}}{\rho^2\varpi}\Big[ 
\left(\frac{\Delta^{1/2}\varpi}{\alpha}
(\Phi_{,r}-\omega G_{,r})\right)_{,r}
\nonumber
\\
&&
+\left( \frac{\varpi}{\alpha\Delta^{1/2}}
(\Phi_{,\theta}-\omega G_{,\theta})\right)_{,\theta} \Big]
=-4\pi\rho_e .
\label{eqnPoisson}
\end{eqnarray}

Once the electromagnetic fields are known, the outgoing energy flux through
 a surface  at a radius $r$ is calculated as follows (see the Appendix):
\begin{equation}
P_{{\rm em}}(r)=- \int (\sqrt{-g} T_{{\rm em} ~t} ^r ) d\theta d\phi 
  =-\frac{1}{2}\int (\Phi,_{\theta} S ) d\theta .
\label{EMpower}
\end{equation}

In the above we have presented a general form of stationary and axially symmetric 
electromagnetic fields, which are described by three functions.
If  the ideal MHD condition  ${\vec E } \cdot {\vec B} =0$ holds,
we have  $\Phi=\Phi(G)$, or 
${\vec \nabla} \Phi=\Omega_{\rm F}(G) {\vec \nabla} G$, where
$\Omega_{\rm F}$ represents the angular velocity of the magnetic field.
Moreover, when the force-free approximation
$ \rho_{e} {\vec E} + {\vec j} \times{\vec B}  =0$ is used, 
the azimuthal component leads to $S=S(G)$.
The electromagnetic fields are described only by the magnetic 
function $G$ (the Grad--Shafranov equation).
The number of differential equations decreases, but there remains the 
problem of specifying their functional relations,
 $\Omega_{\rm F}(G) $ and  $S=S(G)$.
%%%

%%2.2%%%%%%%%%%%%%%%%%%%%%%%%%%%%%%%%%%%%%%%%%%%%%
\subsection{Outgoing energy flux from horizon}
%%%%%%%%%%%%%%%%%%%%%%%%%%%%%%%%%%%%%%%%%%%%%%%%%
To evaluate $P_{{\rm em}}$ in eq.~(\ref{EMpower}) near a black hole 
horizon (at $r_{\rm H} = M+\sqrt{M^2-a^2}$), 
we need the behavior of the functions $\Phi$ and $S$.
For a while, we assume that the ideal MHD 
condition holds, that is, $\Phi ,_{\theta} =\Omega_{\rm F} G,_{\theta}$. 
Near the horizon, the function $S$ is specified by imposing the so-called 
Znajek condition \citep{1978MNRAS.185..833Z},
$  B_{\hat{\phi}}=-E_{\hat{\theta}} $, which is written 
from eqs.~(\ref{Bcomp}),(\ref{Ecomp}) as
\begin{equation}
 S= (\Phi ,_{\theta} -\omega_{\rm H} G,_{\theta})\frac{\varpi}{\rho} , 
\label{ZnajekCND}
\end{equation}
where $\omega_{\rm H} \equiv \omega(r_{\rm H})= a/(2Mr_{\rm H})$ is
the angular velocity of the black hole.
This condition is equivalent to the incoming electromagnetic wave
 \citep{1986bhmp.book.....T}.
By substituting these expressions into eq.~(\ref{EMpower}), the outgoing 
electromagnetic power is
\begin{equation}
P_{{\rm em}}(r_{\rm H})=
-\frac{1}{2}\Omega_{\rm F}(\Omega_{\rm F} -\omega_{\rm H})
\int_{r_{\rm H}}\frac{\varpi}{\rho}(G,_\theta)^2 d\theta.
\label{positiveENG}
\end{equation} 
This shows that the power $P_{\rm em}(r_{\rm H}) $ is positive
when $ 0<\Omega_{\rm F}<\omega_{\rm H}$.
This is the outgoing energy flux from a rotating black hole
in the BZ process.
A remaining problem is how the angular velocity $ \Omega_{\rm F}$ of the magnetic
field lines or equivalently the longitudinal electric potential $\Phi $ is determined.

%%%2.3%%%%%%%%%%%%%%%%%%%%%
\subsection{Fluid}
%%%%%%%%%%%%%%%%%%%%%%%%%%%
  We adopt a treatment in which the plasma is modeled as 
a two-component fluid. Each component, consisting of positively 
or negatively charged particles, is described by a number density
$n_{\pm }$ and velocity $\vec{v}_{\pm }$,
which denote the values measured by a ZAMO. 
The proper density $n_{\pm }^{\ast }$ measured in the fluid rest frame is 
related to $n_{\pm }$ and  a Lorentz factor 
$\gamma _{\pm }=(1-(v_{\pm}/c)^{2})^{-1/2}$ 
by $n_{\pm }^{\ast }=( n /\gamma)_{\pm} $,
where an abbreviation
 $( n /\gamma)_{\pm} \equiv n_{\pm }/\gamma _{\pm} $
is introduced.
We assume that the positive particle has mass $m$ and charge $e$, 
while the negative one has mass $m$ and charge $-e$. 
The charge density and electric current are given in terms of 
$n_{\pm }$ and $\vec{v}_{\pm }$ as
\begin{eqnarray}
\rho _{e} &=&e(n_{+}-n_{-}), 
   \label{dfn.charge} \\
\vec{j} &=&e(n_{+}\vec{v}_{+}-n_{-}\vec{v}_{-}).
   \label{dfn.current}
\end{eqnarray}
The continuity equation for each component in the 
axisymmetric stationary condition is 
\begin{equation}
0=\vec{\nabla}\cdot (\alpha n^{*} \gamma \vec{v}_{p} )_{\pm } 
=\vec{\nabla}\cdot (\alpha n \vec{v}_{p})_{\pm } ,
  \label{eqn.cont}
\end{equation}
where the factor $ \alpha $ comes from the relation between the
determinant of 4-dimensional spacetime metrics and 
that of 3-dimensional space
$ \sqrt{-g_{4}} = \alpha \sqrt{g_{3}}$.
This equation is satisfied by introducing a stream function 
$F_{\pm}(r,\theta )$ as 
\begin{equation}
\alpha (n \vec{v}_{p})_{\pm} =
\frac{1}{\varpi}\vec{\nabla}F_{\pm }\times 
\vec{e}_{\hat{\phi}}.
  \label{dfn.fluid}
\end{equation}
From this definition, the number density is solved as 
\begin{eqnarray}
n_{\pm } &=& (\alpha \varpi)^{-1} (
|\nabla F| (v_{\hat r}^{2}+v_{\hat \theta }^{2})^{-1/2} )_{\pm}
\nonumber
\\
&=& (\alpha \varpi)^{-1} (|\nabla F| 
 (1 -\gamma^{-2} -v_{\hat \phi }^{2} )^{-1/2} )_{\pm} .
  \label{dfn.density}
\end{eqnarray}
From eqs.~(\ref{dfn.current}) and (\ref{dfn.fluid}), 
the current flow function $S$ of
eq.~(\ref{enqjcpol}) can be given by 
\begin{equation}
S=4\pi e (F_{+}-F_{-}).
\label{relSandFpm} 
\end{equation}

The law of momentum conservation in a stationary 
axially symmetric state \citep{1998MNRAS.294..673K} is
\begin{equation}
\frac{1}{\alpha} \nabla _{j} ( \alpha T^{j} _{\pm ~i} )
= 
\rho_{\rm m  \pm} g_{i} + H_{ij} S_{\rm m \pm} ^{j}
\pm e n^{*} _{\pm} \gamma _{\pm }
(\vec{E}+\vec{v}_{\pm } \times \vec{B})_{i} 
~{\pm }(R _{\rm col})_{i},
\label{motion0.eqn}
\end{equation}
where the first term denotes gravitational acceleration
with ${\vec g} \equiv -\vec{\nabla} \ln \alpha $,  
the second is a gravito-magnetic term with 
$ H_{ij} \equiv \alpha^{-1} \nabla _{i} \beta_{j}$, 
the third is the electromagnetic force, and 
the last is the collision term.
Here we consider the cold limit,
so that thermal pressure is ignored, and
the stress tensor is 
$T_{\pm} ^{ij} = (mn^{*}\gamma ^2 v^iv^j)_{\pm}$,
energy density $\rho _{\rm m \pm} = (mn^{*}\gamma ^2 )_{\pm}$,
and momentum flux 
$S_{\rm m \pm} ^{i} = (mn^{*}\gamma ^2 v^i)_{\pm}$.
Moreover, we neglect the collision term.
As \citet{1998MNRAS.294..673K} discussed, near the black hole horizon, 
the electron collision time becomes longer than the dynamical timescale,
so the collisionless approximation may be valid under certain 
conditions.
Using the collision rate $\nu_c$~\citep{1962pfig.book.....S}
and free-fall timescale $t_{\rm ff}$, 
the product is $\nu_c t_{\rm ff} \sim 10^{-3}$,
where the electron number density 
$n_{e} \sim 10^{16} ({\dot M}/10^{-2}{\dot M}_{\rm E})(M/ M_{\odot})^{-1}$
${\rm cm}^{-3}$,
estimated from the accretion rate, and thermal velocity at $T=10^{12}$K are 
used. Collisions are ignored in the dilute approximation. 
Thus, the equation of motion (\ref{motion0.eqn})
for each component through the global electromagnetic and 
gravitational forces is reduced to
\begin{equation}
[(\vec{\nabla}\times \gamma \vec{v})\times \vec{v}
+\vec{\nabla}\gamma ]_{\pm ~i}
= 
\gamma _{\pm } g_{i} + H_{ij} (\gamma v^{j} )_{\pm } 
\pm \frac{e}{m}
\left[ \vec{E}+\vec{v}_{\pm }\times \vec{B}\right]_{i} ,
 \label{motion.eqn}
\end{equation}
where the left-hand side is written in the vector form.
From eq.~(\ref{motion.eqn}), it is clear that there are two 
conserved quantities along each stream line, namely,
the generalized angular momentum $J_{\pm }$
and the Bernoulli integral $K_{\pm }$, which are
equivalent to $ u_\phi$ and $u_t $ for each fluid component: 
\begin{equation}
J_{\pm }= (\varpi \gamma v_{\hat \phi} )_{\pm }
\pm \frac{e}{m}G ,
   \label{const.Ang}
\end{equation}
\begin{equation}
K_{\pm }= ( \omega \varpi\gamma v_{\hat \phi} )_{\pm } +
\alpha \gamma _{\pm }\pm \frac{e}{m}\Phi .
   \label{const.Bern}
\end{equation}
These quantities depend on only the stream functions $F_{\pm }$, 
and the spatial distributions are therefore determined by $F_{\pm }$, 
which is specified at the injection point in our model.
The component of eq.~(\ref{motion.eqn}) perpendicular
to the stream lines gives

%%%
\begin{eqnarray}
&&
\alpha \varpi ^2 {\vec \nabla } \cdot
\left( \frac{\gamma  _{\pm } }{ \alpha n_{\pm } \varpi ^2} 
{\vec \nabla } F_{\pm} \right) 
\nonumber
\\
&&~~
=
\alpha \varpi^{2} n_{\pm }
\left( K_{\pm }^{\prime }
 -\left( v^{\pm} _{\hat \phi} +w_{0} \right)
 \frac{\alpha}{\varpi} J_{\pm}^{\prime}\right)
 \pm \frac{e}{m}S,
  \label{eqn.trans1}
\end{eqnarray}
%{equation}
where $w_{0} =\omega \varpi /\alpha$, and 
$J_{\pm }^{\prime }$ and $K_{\pm }^{\prime }$ are derivatives of 
$J_{\pm }$ and $K_{\pm }$ with respect to $F_{\pm }$.

We have thus obtained a system of equations that
govern the electromagnetic field structure and plasma flows.
Four partial differential equations (\ref{eqnBioG}), (\ref{eqnPoisson}), 
(\ref{eqn.trans1}) should be solved with two 
integrals (\ref{const.Ang}), (\ref{const.Bern})
and the number density (\ref{dfn.density}) derived by the stream functions.
These are reduced to those obtained for pulsar electrodynamics
in a flat spacetime \citep{2009MNRAS.398..271K} by setting $M=a=0$.
These equations for $G$, $\Phi $, and $F_{\pm }$ are interdependent in a nonlinear manner, so iterative methods are needed 
to self-consistently solve a set of these equations.
For example, assume that functions $G$, $\Psi $, 
and $F_{\pm }$ are known.
The azimuthal velocity $v_{\hat \phi}$ and Lorentz factor $\gamma$
are determined by the integrals in Eqs.~(\ref{const.Ang}) and (\ref{const.Bern}).
The number density is calculated from eq.~(\ref{dfn.density}).
Thus, the source terms, namely, the toroidal current $j_{\hat \phi} $ in eq.~(\ref{eqnBioG}) and 
the charge density $\rho_{e}$ in eq.~(\ref{eqnPoisson}), are calculated from 
the fluid quantities of both species. The source terms and complicated 
coefficients in the equation of $F_\pm$ are also calculated.
A new set of functions is solved for these source terms with 
appropriate boundary conditions. This procedure is repeated 
until convergence.

%(3)%%%%%%%%%%%%%%%%%%%%%%%%%%%%%%%%%%%%%%%%%%%
\section{Model description}
%%%%%%%%%%%%%%%%%%%%%%%%%%%%%%%%%%%%%%%%%%%%%%%
%%3.1%%%%%%%%%%%%%%%%%%%%%%%%
\subsection{Parameters and normalization}
%%%%%%%%%%%%%%%%%%%%%%%%%%%%
We now describe a general framework for determining the
electromagnetic fields and charged flows described in the previous section.
Here, we discuss physical parameters involved in our system.

There are two independent parameters.
One is a dimensionless gyrofrequency,
$ \chi \equiv \omega_{B} M = e B_{0} M/m$, 
where $ B_{0} $ is a typical magnetic field strength.
Associated with $ B_{0} $ is a characteristic number
density $ n_{c} \equiv B_{0}/(4\pi e M)$. The number density is 
reduced to the so-called Goldreich-Julian density if the 
timescale $2M$ is replaced by the stellar angular 
velocity $\Omega_s ^{-1}$. 
The actual number density is normalized by $ \lambda  n_{c}$, where
$ \lambda $ represents the multiplicity of pair plasma.
We can express other physically meaningful quantities
using these dimensionless parameters $\chi$ and $\lambda$.
The normalized plasma frequency $\kappa$ with number 
density $\lambda  n_{c}$ is given by $\kappa ^2 \equiv \omega_{p} ^2 M^2$
$= 4\pi e^2 (\lambda  n_{c})M^2/m =\lambda \chi $.
The ratio of $\lambda$ to  $\chi $ is written as 
$k \equiv \lambda / \chi = (1/4) \times (2m \lambda n_{c})/(B_{0} ^2 /8 \pi)$,
and represents the ratio of the rest mass energy density of pairs 
to the electromagnetic energy density. 
These numbers $\chi $ and $\lambda $ are very large in astrophysical situations.
A typical value of $\chi$ is 
$ 10^{13} (B_{0}/{\rm kG})(M/10^8 M_{\odot})$,
relevant to active galactic nuclei powers $ \propto (MB_0) ^{2}$.
The amount of pair plasma is unclear, but we here estimate it from the accretion 
rate ${\dot M}$. Using the electron number density 
$n_{e} \sim 10^8 ({\dot M}/10^{-2}{\dot M}_E)(M/10^8 M_{\odot})^{-1}$
${\rm cm}^{-3}$ near the horizon, we have 
$\kappa^2 = 10^{22} ({\dot M}/10^{-2}{\dot M}_E)(M/10^8 M_{\odot})$.
The other parameters are calculated as
$ \lambda =\kappa^2/\chi \sim 10^{9}$ and $k=\kappa^2/\chi^2 \sim 10^{-4}$.
See also \citet{2010PhyU...53.1199B} for estimates for 
microquasars and gamma ray bursts,
for which $ \lambda \sim 10^{10}$--$10^{14}$.
It is true that $\lambda , \chi, \kappa^2 \gg 1 $, but
these values should be regarded as an order estimate with large uncertainties.
In particular, the ratio $ k =\lambda/\chi $ may drastically change 
in the cases of
matter-dominated or magnetically dominated flows.
Indeed, the activation condition of the BZ mechanism in the MHD flows
is approximately given by $k < 1 $, 
for which the Alfven speed exceeds the free-fall velocity
at the ergosphere\citep{2009MNRAS.397.1153K}.
Magnetization parameter corresponds to $1/k $.

%%
%%%%%%%%%%%%%%%%%%%%%%%%%%%%%%%%%%%%%%%%%%%%%%%
%%%%% FIG0 %%%%%
 \begin{figure}
\centering
\includegraphics[scale=1.0]{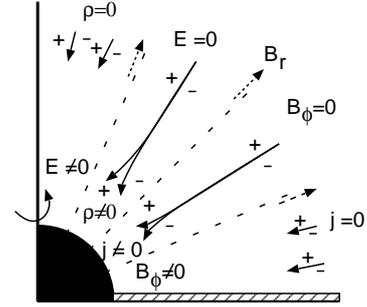}
\caption{ A schematic illustration for electromagnetic fields and 
plasma inflows around a rotating black hole.
The magnetic field is radial without toroidal component, and 
electric field vanishes in the outer region.
Black hole rotation affects the plasma flow, and a new
electromagnetic structure is induced in the inner region.
There is a current sheet on an equator 
to support the split-monople magnetic field. 
}
\label{fig0}
\end{figure}
%%%%%%%%%%%%%%%%%%%%%%%%%%%%%%%%%%%%%%%%%%%%%%%

%%
We provide an explicit model of electromagnetic fields and plasma 
inflows in a hemisphere $ (0 \le \theta \le \pi/2)$,
depicted schematically in Fig.~\ref{fig0}.
We assume that the electromagnetic fields at $r_{{\rm out}} \gg M$
are described by ${\vec B} = B_0 (M/r)^2 {\vec e}_{\hat r}$, $B_{\hat{\phi}} =0$
and ${\vec E} =0$, where $B_{0}$ is a constant representing 
magnetic field strength.
This condition differs from the wind solution by 
\citet{1973ApJ...180L.133M}
in which toroidal magnetic field and electric fields are given by
$ B _{\hat{\phi}} =$$ E_{\hat{\theta}} =$
$- \Omega_{\rm F} B_0 M^2 \sin \theta /r $.
Our concern is how the parameter $\Omega_{\rm F}$ is determined, 
so that the condition $\Omega_{\rm F}=0$ is used at $r_{{\rm out}} $.
Such a split monopole magnetic field may be formed by
strong currents on an equatorial disk,
by which upper and lower hemispheres are detached.
The electromagnetic fields are obtained by the derivatives of  
${\bar G} \equiv G/(B_{0}M^2)= (1-\cos\theta)$,
${\bar S} \equiv S/(B_{0}M) =0$ and ${\bar \Phi} \equiv \Phi/(B_{0}M) =0$,
where ${\bar G} $, ${\bar S} $ and ${\bar \Phi} $ are
normalized dimensionless quantities.

As for the pair plasma, we assume  neutral flow falling along the 
radial magnetic fields at $r_{{\rm out}} \gg M$, that is,
$v^{\pm} _{\hat\phi} \to 0$ and $\alpha \gamma_{\pm} \to 1$.
The stream functions $F _{\pm } $ of both components
should coincide there, since $F _{+}-F _{-}=S/(4\pi e)=0$
in eq.~(\ref{relSandFpm}).
Like the magnetic function $G$, both functions are chosen to be
radial, $F _{\pm } = -\lambda n_{c} M^2(1-\cos\theta)$,
where $ n_{c} \equiv B_{0}/(4\pi e M)$ is a characteristic number density 
and the minus sign denotes inflow $v^{\pm} _{\hat{r}} <0$. 
We introduce dimensionless stream functions
${\bar F}_{\pm} =F_{\pm}/(\lambda n_{c}M^2)$ and the dimensionless number 
density ${\bar n}_{\pm} =n_{\pm }/(\lambda n_{c})$.
The relation (\ref{relSandFpm}) between ${\bar S}$ and ${\bar F}_{\pm}$
becomes ${\bar S} =\lambda ({\bar F}_{+} - {\bar F}_{-})$.

Under these conditions at $r_{{\rm out}} $, the integrals $J_{\pm}$ 
and $K_{\pm}$ in eqs.~(\ref{const.Ang}) and (\ref{const.Bern})
are explicitly given by
\begin{equation}
 J_{\pm } =\mp \frac{\chi}{\lambda n_{c}M} F_{\pm }
= \mp \chi M {\bar F}_{\pm },
%   \label{eqn.Jmodel}
~~
K_{\pm} = \alpha \gamma_{\pm}  = 1,
   \label{eqn.JKmodel}
\end{equation}
where the Lorentz factor at $r_{{\rm out}} (\gg M)$ is for simplicity 
chosen as $\gamma_{\pm} =\alpha^{-1} (\approx 1)$.
Once $J_{\pm}$ is specified, the azimuthal velocity $v^{\pm} _{{\hat \phi}}$
can be solved at any point from eq.~(\ref{const.Ang}) as 
\begin{equation}
v^{\pm} _{{\hat \phi}} = \mp \frac{\chi M}{\varpi \gamma _{\pm} }
( {\bar F}_{\pm } + {\bar G} ).
\label{eqn.v3slv}
\end{equation}
The Lorentz factor $\gamma _{\pm}$ can also be solved 
from eq.~(\ref{const.Bern}) as 
\begin{equation}
\alpha \gamma _{\pm } = 1 \pm \chi
\left[ \omega M ( {\bar F}_{\pm } + {\bar G} )
- {\bar \Phi} \right] .
   \label{eqn.gmslv}
\end{equation}
Since we have $K_{\pm} ^\prime =0 $ and  
$J_{\pm} ^\prime =\mp \chi / (\lambda n_{c}M) $,
eq.~(\ref{eqn.trans1}) is reduced to
\begin{equation}
\alpha \varpi ^2 {\vec \nabla } \cdot
\left( \frac{M^2 \gamma  _{\pm } }{ \alpha {\bar n}_{\pm } \varpi ^2} 
{\vec \nabla }{\bar F}_{\pm} \right) 
= \pm \chi
\left[ \frac{\alpha ^2 \varpi {\bar n}_{\pm } }{M} 
 \left(w_{0} + v^{\pm} _{\hat\phi} \right)  + {\bar S}
\right],
\label{eqn.transnn}
\end{equation}
where $ w_{0} = \omega \varpi/\alpha $.

In this model, we have imposed the conditions $S=0$ and $\Phi =0$
at $r_{\rm out}$, so the electromagnetic Poynting power (\ref{EMpower})
is zero ($P_{\rm em}(r_{\rm out})=0$).
If the ideal MHD condition holds in $r_{\rm H} \le r \le r_{\rm out}$, 
then we have $ \Phi =0$ everywhere, including 
the black hole horizon as the asymptotic limit,
since the constant $\Omega_{\rm F}$ along any magnetic field line 
is zero. Consequently, no electromagnetic power is produced.
Our concern is how power is produced in the presence of black hole 
spin. For this purpose, we have to consider
non-ideal MHD effects.

%%3.2%%%%%%%%%%%%%%%%%%%%%%%%%%
\subsection{Spherical flow}
%%%%%%%%%%%%%%%%%%%%%%%%%%%%%%%
Here, we discuss the structure of the electromagnetic field and 
plasma flows in Schwarzschild spacetime.
An analytic solution is given in terms of dimensionless functions as
\begin{equation}
 {\bar G} = -{\bar F}_{\pm} = 1-\cos\theta,~
 {\bar \Phi} ={\bar S}=0 .
\label{eqn.sphGFP}
\end{equation}
The electromagnetic fields are explicitly written as
\begin{equation}
[B_{\hat{r}},B_{\hat{\theta}}, B_{\hat{\phi}}]
=\left[\frac{B_{0} M^2}{r^2},0, 0\right],
~~
[E_{\hat{r}},E_{\hat{\theta}}, E_{\hat{\phi}}]
=\left [0,0,0\right].
\end{equation}
The flow velocity, its Lorentz factor, and 
number density of the flow are given by 
\begin{eqnarray}
&&
[v^{\pm} _{\hat{r}},v^{\pm} _{\hat{\theta}}, v^{\pm} _{\hat{\phi}}]
=\left[-\left(\frac{2 M}{r}\right)^{1/2},0, 0\right],
\nonumber
\\
&&
\gamma_{\pm} =
\gamma_{0} = (1-(v_{\pm})^2)^{-1/2} = \alpha^{-1},
\nonumber
\\
&&
{\bar n}_{0} \equiv \frac{n_{\pm} }{\lambda n_{c}}= 
\frac{1}{\sqrt{2}}\frac{M^{3/2}}{\sqrt{r^2(r-2M)}}.
\label{eqn.sphVN}
\end{eqnarray}
The number density $n_{\pm}$ appears to diverge at the horizon $ r \to 2M$,
but the proper density $n_{* \pm}=n_{\pm} /\gamma _{0}$ is finite everywhere, as
$n_{\pm} /(\lambda n_{c} \gamma _{0})=$ $M^{3/2}/(\sqrt{2} r^{3/2})$.
The proper density and the magnetic field strength at $r=2M$ 
are  $n_{* \pm}=$ $ \lambda n_{c}/4 $, and 
$B_{\hat r} =B_{0}/4 \equiv B_{\rm n} $.
Thus, $B_0$ and $\lambda n_{c}$ are reasonable values
near the black hole horizon, estimated in the previous section. 

This is a spherically symmetric solution, so that the conditions imposed 
at radius $r_{\rm out}$ are retained everywhere. 
In particular, the electromagnetic power is zero everywhere.
The energy flow $P_{0}$ of matter across a sphere with radius $r$ 
is obtained by twice the value in a hemisphere 
($0 \le \theta \le \pi/2$):
\begin{equation}
P_{0} = -2\times  4\pi \lambda n_{c} m M^2
= -\frac{ 2 \lambda }{\chi} B_{0}^2 M^2.
\label{MATpower0}
\end{equation}
Here, a minus sign denotes inflow, and 
the factor 2 comes from summing the contribution of
positively and negatively charged fluids (see the Appendix).
%%

%%3.3%%%%%%%%%%%%%%%%%%%%%%%%%%%%%%
\subsection{Effect of slow rotation}
%%%%%%%%%%%%%%%%%%%%%%%%%%%%%%%%%%%
We examine the effect of black hole spin on the spherical flow
given by eqs.~(\ref{eqn.sphGFP})--(\ref{eqn.sphVN}).
It is rather difficult numerical work to 
obtain consistent solutions for $ G, \Phi$, and $ F_{\pm} $ 
in eqs.~(\ref{eqnBioG}), (\ref{eqnPoisson}), and (\ref{eqn.transnn}), as
they consist of nonlinearly coupled partial differential equations.
We here consider the rotation as a small parameter, 
and expand these functions as, for example, 
$ F_{\pm}= \lambda n_{c}M^2({\bar F} + \delta F_{\pm})$.
We limit our consideration to the first-order effect.
The Poynting flux given by eq.~(\ref{EMpower}) is a product 
of $\Phi_{,\theta}$ and $S$. Both are zero for $a=0$, but are modified 
by the first-order rotational effect.
Poynting flux is thus produced within this approximation level.
Within the first-order effect of the black hole spin, the only difference
from the Schwarzschild metrics is the function $\omega $, 
which is approximated as $\omega = 2M^2 a_{*} r^{-3}$, where
$a_{*} =a/M$ is a dimensionless Kerr parameter.
We assume $a_{*}  >0 $, which determines the direction of 
perturbed vectors.
In this section, the symbols $\alpha  $ and  $\varpi $
denote those for $a_{*}=0$, that is, $\alpha^2 = 1 -2M/r $
and $\varpi  =r \sin \theta $.

We next develop the perturbation equations.
We first note that the rotational effect 
in eq.~(\ref{eqn.transnn}) is the term $w_{0} =\alpha^{-1}\omega \varpi$ 
in only the lowest approximation, and that the term in the equation of the stream functions $F_{\pm }$ works 
in an opposite direction with respect to the fluid species.
We therefore consider only a class of perturbations
$ \delta F_{+}= -\delta F_{-} $.
From the perturbation of eq.~(\ref{eqn.v3slv}),
$\delta v^{\pm} _{\hat\phi} $ is expressed
by $\delta F^{\pm }$ and $\delta G$,
but after careful consideration we conclude that $\delta G =0 $ and 
$\delta v^{+} _{\hat\phi} =  \delta v^{-} _{\hat\phi} $.
We also  have $\delta G =0 $ in the perturbation of 
eq.~(\ref{eqnBioG}), consistent with 
$0=\delta j_{\hat \phi} $
$\propto (\delta v^{+} _{\hat\phi} -\delta v^{-} _{\hat\phi}) $.
From now on, we will use a function 
$ \delta F \equiv \delta F_{+}=- \delta F_{-}$, and
$\delta v^{\pm} _{\hat\phi}$ is given by
%%%
\begin{equation}
\delta v^{\pm} _{\hat\phi} =
 -\chi M(\varpi \gamma_{0})^{-1}\delta F.
\label{eqn.deltv3}
 \end{equation}
Under these conditions the perturbation of eq.~(\ref{eqn.gmslv})
is reduced to
\begin{equation}
 \delta \gamma ^{\pm}  =\mp \chi \alpha ^{-1} \delta \Phi
=\mp\chi \gamma_{0} \delta \Phi.
\end{equation}

The acceleration, which is opposite that of the
fluid species, originates from the perturbation of electric 
potential.
The perturbation of number density, eq.~(\ref{dfn.density}),
is also opposite to the species and is given by
\begin{equation}
\frac{ \delta n_{\pm} }{ {\bar n}_{0}}=
\pm \left[ -\frac{1}{\sin \theta} \delta F_{, \theta }
+ \chi ( \gamma_{0} ^2 -1)^{-1} \delta \Phi \right].
\label{eqn.denpert}
\end{equation}

Finally, we consider the perturbation equations
for $\delta \Phi $  and $\delta F$. 
General forms expanded with the Legendre polynomials 
$P_l (\theta)$ are given by
$\delta \Phi =\sum h_{l} (r) P_l (\theta)$
and  $\delta F =-\sum (l+1)^{-1} p_{l} (r) P_{l, \theta} (\theta)\sin \theta$.
The slow rotation corresponds to the dipole perturbation with $l=1$, so that 
the components with $l \ne 1$ are decoupled with the rotational 
perturbation. Therefore we have  
$ \delta \Phi = h(r) \cos \theta $ and 
$ \delta F = (1/2) p(r) \sin^2 \theta $
\footnote{
The form $ \delta F $ allows stream lines to hit on the equator, 
and therefore poloidal current may go into and out it.
This point will be discussed later.
}.
Substituting their forms into the perturbation equations of 
Eqs.~(\ref{eqnPoisson}) and (\ref{eqn.transnn}),
after some manipulations we derive the following:
% final % %% p & h 
%
\begin{equation}
\frac{\alpha^2}{s^{2}}\frac{d}{ds}
\left(s^{2} \frac{d h}{ds} \right)
= - \left[ \frac{ \kappa^2 \alpha^2}{\sqrt{2s}}-\frac{2}{s^{2}}\right] h
+  \frac{\sqrt{2}\lambda}{s^{3/2}} p + \frac{4 a_{*}}{ s^{5}},
\label{eqn2x1}
\end{equation}
\begin{equation}
\alpha^2 \frac{d}{ds}\left( \frac{s^{3/2}}{\alpha^2}
\frac{d p}{ds} \right)
= \left[ \sqrt{2} \kappa^2 - \frac{\chi^2 \alpha^{2}}{2 s^{3/2}} \right]p
+ \chi s^{1/2} h + \frac{2\chi a_{*}}{ s^{5/2}}.
\label{eqn2x2}
\end{equation}
Here, we use normalized radial coordinate $s \equiv r/M$. We thus have 
a coupled set of second-order ordinary differential equations.
There are very large numbers involved in the first terms on the right-hand side, namely
the squares of the plasma frequency $\kappa$ and gyrofrequency $\chi$.
The last terms represent the effect of black hole spin $a_*$.  
Our concern is the range $\chi \gg 1 $ 
and  $\kappa^2 = \lambda \chi \gg 1 $, but
$k = \lambda /\chi$ is not so large.
Replacing $\kappa^2 =k \chi^2$ in eqs.~(\ref{eqn2x1}) and (\ref{eqn2x2}), 
we neglect higher order terms of $\chi^{-n} ~(n > 2)$
except the derivative terms.
Thus, eqs.~(\ref{eqn2x1}) and (\ref{eqn2x2}) are approximated 
in forms suitable for WKB analysis:
%%%  WKB %%%
\begin{equation}
\left[ \chi^{-2} \frac{d^2}{ds_* ^2}
+U \right] \left( \frac{s h }{\alpha} \right) 
 -\chi^{-1} \sqrt{2}k A
\left(\frac{s^{3/4} p}{\alpha^2} \right) = 0,
\label{eqnWKBUU}
\end{equation}
\begin{equation}
\left[
\chi^{-2} \frac{d^2}{ds_* ^2}
-V 
\right]\left(\frac{s^{3/4} p}{\alpha^2} \right)
 -\chi^{-1} A  \left( \frac{s h }{\alpha} \right) 
 = \chi^{-1} J_s.
\label{eqnWKBVV}
\end{equation}
%%%
Here, $s_*$ in eqs.~(\ref{eqnWKBUU})--(\ref{eqnWKBVV}) denotes 
tortoise coordinate $s_{*} \equiv r_{*}/M$$=(r/M)+ 2\log(r/2M-1)$
satisfying $ds_{*}/ds = \alpha ^{-2}$.
Other terms in eqs.~(\ref{eqnWKBUU})--(\ref{eqnWKBVV}) are 
\begin{eqnarray}
&&
U =(k/\sqrt{2}) \alpha^4 s^{-1/2}, 
~~
V =\left( \sqrt{2}k -\frac{1}{2} \alpha^2 s^{-3/2} \right)
  \alpha^4 s^{-3/2} ,
\nonumber
\\
&&
A =\alpha^3 s^{-5/4} ,
~~
J_s =2 a_{*} \alpha^2 s^{-13/4} .
\end{eqnarray}
The system of Eqs.~(\ref{eqnWKBUU})--(\ref{eqnWKBVV}) is rather
simplified, since the source term in eq.~(\ref{eqnWKBUU}), the
higher order term $\sim \chi^{-2}$, is neglected,
and the potential terms $U$ and $V$ involve a parameter
$k= \lambda/\chi $.

%%3.4%%%%%%%%%%%%%%%%%%%%%%%%%%
\subsection{WKB solutions}
%%%%%%%%%%%%%%%%%%%%%%%%%%%%%%%
We first consider the homogeneous solution of 
eqs.~(\ref{eqnWKBUU})--(\ref{eqnWKBVV}).
We seek an approximate solution of the WKB form
$ p \propto \exp( \chi W(s) )$ and $ h \propto \exp( \chi W(s) )$,
where $\chi ^{-1} (\ll 1)$ is a small WKB parameter.
Substituting them in, we find the 
leading-order solutions correct to order $ \chi ^{-1}$.
The four independent solutions (two pairs) given below are denoted 
by $h^{\pm} _{n}, p^{\pm} _{n}$. 
Two types are clearly distinguished among the three points:
(1) oscillatory or growing/decaying behaviors,
(2) the relative ordering between $h$ and $p$
and (3) their relative sign.
A pair of type I solutions is given by
\begin{eqnarray}
&&
h^{\pm} _{{\rm I}}  = \alpha s^{-1} U^{-1/4}
 \exp(\pm i \chi \int^{s_*} U^{1/2} ds_* ^\prime )
\nonumber
\\
&& ~~
=  2^{1/8}k^{-1/4} s^{-7/8} \exp(\pm i \kappa_1  s^{3/4} ),
\label{eqn.typuu1}
\end{eqnarray}
\begin{equation}
p^{\pm} _{{\rm I}} = -2 \chi^{-1} s^{1/2}
 \left[ 2^{1/2} k (s+2) - \alpha ^2 s^{-3/2}
\right]^{-1} h^{\pm} _{{\rm I}}, 
\label{eqn.typuu}
\end{equation}
where $\kappa_1 = (2^{7/4}/3)\kappa $ and an overall constant 
from the integral is adjusted in eq.~(\ref{eqn.typuu1}).
The solution represents spatial oscillation of the plasma,
whose wavelength is $\sim \kappa^{-1} M$.
The solution satisfies the relations $p^{\pm} _{{\rm I}} $
$\sim \chi ^{-1}\times h^{\pm} _{{\rm I}}$
$\ll h^{\pm} _{{\rm I}}$ in the large $\chi$ limit,
and $ h^{\pm} _{{\rm I}} p^{\pm} _{{\rm I}} <0$ if
the value in square brackets in eq.~(\ref{eqn.typuu}) is positive.
The last condition is satisfied 
when  $k = \lambda/\chi  > 9.1 \times 10^{-3}$.
%%--
The functions $ h^{\pm} _{{\rm I}}$ and $ p^{\pm} _{{\rm I}} $
are regular toward the black hole horizon $ \alpha  \to 0,  r \to 2M$.

%%2%%
Another pair of type II solutions is
%\begin{eqnarray}
\begin{equation}
%&&
p^{\pm} _{{\rm II}} = 
\alpha^{2} s^{-3/4} V^{-1/4}  
\exp( \pm \chi \int^{s_*} V^{1/2} ds_* ^\prime ) ,
\label{eqn.typv0}
\end{equation}
%\\&&
\begin{equation}
h^{\pm} _{{\rm II}}  = 2^{3/2} \chi^{-1} k \alpha ^{-2} \left[
 2^{1/2} k (s+2) - \alpha ^2 s^{-3/2} \right]^{-1}
p^{\pm} _{{\rm II}} .
\label{eqn.typvv}
\end{equation}
%\end{eqnarray}
%
Here, we assume that $V$ is positive everywhere.
This condition is satisfied for the parameter
$k = \lambda/\chi  >k_c $, $k_c \approx 2.3\times 10^{-2}$. 
When  $k < k_c$ the potential $V$ becomes negative in some
range $r_1< r < r_2$, and the function in eq.~(\ref{eqn.typv0}) becomes 
oscillatory there. A whole solution is obtained
by matching functions at $r_1 $ and $r_2 $.
We expect that such a solution is possible 
for only a discrete value of $k$, namely, an eigenvalue,
and requires more careful treatment.
In the following, our consideration is limited to the case $k >k_c $.
Note that values in the square brackets in eqs.~(\ref{eqn.typuu}) and (\ref{eqn.typvv}) are positive for that case as well.

For later convenience, we approximate the integral in the 
exponential in eq.~(\ref{eqn.typv0}).
With a constant $\kappa_{2} = 2^{9/4}\kappa$,  
eq.~(\ref{eqn.typv0}) is reduced to
\begin{equation}
p^{\pm} _{{\rm II}} \approx  2^{-1/8} k^{-1/4}\alpha s^{-3/8}
\exp( \pm  \kappa_{2} s^{1/4} ).
\end{equation}
We numerically verified that the approximation is good so long
as $\kappa \gg 1$ and $k  > k_c$.

We next discuss properties of the type II solutions 
in eqs.~(\ref{eqn.typv0}) and (\ref{eqn.typvv}).
These are exponentially growing/decaying functions with
a relation $h^{\pm} _{{\rm II}} $
$\sim \chi ^{-1}\times p^{\pm} _{{\rm II}}$
$\ll p^{\pm} _{{\rm II}} $ in the large $\chi$ limit.
This ordering is opposite to that in
$ h^{\pm} _{{\rm I}} $ and $ p^{\pm} _{{\rm I}} $.
As discussed in the next subsection, the sign of $ h^{\pm} _{{\rm II}} p^{\pm} _{{\rm II}} >0$
is critical for outgoing energy power.
The function $ p^{\pm} _{{\rm II}} \propto \alpha $ goes to zero at the 
horizon $\alpha  \to 0 $, while $ h^{\pm} _{{\rm II}}$ in general
diverges as $h^{\pm} _{{\rm II}} \sim  \alpha^{-1} $.
The divergence in $h_{{\rm II}} $ will be eliminated by appropriate 
combination of $ p^{\pm} _{{\rm II}} $, as discussed below.

  A general solution of eqs.~(\ref{eqnWKBUU}) and (\ref{eqnWKBVV})
without the source term $J_s$ is expressed by a linear combination 
of four functions as $ h = \sum c^{\pm} _{n} h^{\pm} _{n}(s)$, 
and $ p = \sum c^{\pm} _{n} p^{\pm} _{n}(s)$.
The solution of the inhomogeneous equation
is obtained by varying the coefficients $ c^{\pm} _{n} $ as
 $ h = \sum c^{\pm} _{n} (s) h^{\pm} _{n}(s)$, 
and $ p = \sum c^{\pm} _{n} (s) p^{\pm} _{n}(s)$.
We put these forms into eqs.~(\ref{eqnWKBUU}) and (\ref{eqnWKBVV}),
and find that $ d c^{\pm} _{{\rm I}}/ds_*  \propto  \chi^{-1} $
and that the functions $ c^{\pm} _{{\rm II}} $ 
satisfy the following equation of order $\chi^{0}$:
\begin{equation}
\frac{d c^{\pm} _{{\rm II}} }{ds_*}  =
\pm \frac{1}{2} \frac{J_s}{V^{1/4}} 
\exp( \mp \chi \int ^{s_*} V^{1/2} d s_* ^\prime ) .
\label{c2dfeq}
\end{equation}
We neglect corrections to the type I solution, and consider only the effect
of source term $J_s$ on $ c^{\pm} _{{\rm II}} $. 
By integrating eq.~(\ref{c2dfeq}), a particular solution of the 
inhomogeneous equation is written in a concise form as
%\begin{eqnarray}
\begin{equation}
%&&
 p^{\rm S}_{{\rm II}} = - \frac{a_* \alpha^{2} }{s^{3/4} V^{1/4}}  
 \int _{\rm in} ^{\rm out} 
\frac{\alpha^{2}}{ \xi^{13/4}  V^{1/4}} 
\exp( -\chi |\int_{\xi } ^{s} V^{1/2} \alpha^{-2} ds ^\prime | ) d\xi
\end{equation}
%\\
%&&
\begin{equation}
 ~~ \approx 
-\frac{a_{*} \alpha }{2^{1/4}k^{1/2} s^{3/8} }
 \int _{\rm in} ^{\rm out} 
\alpha^{-1} \xi^{-23/8} \exp( -\kappa_2 | s^{1/4} -\xi^{1/4} | ) d\xi .
\label{eqn.wkb.ps}
\end{equation}
%\end{eqnarray}
This is a method to solve inhomogenous equations in terms of 
a Green function constructed by the WKB approximation
\citep[see, e.g.,][]{1999amms.book.....B}.
  The solution $p$ of eqs.~(\ref{eqnWKBUU}) and (\ref{eqnWKBVV})
is in general a sum of $p^{S} _{{\rm II}}$ and $p^{\pm} _{{\rm II}}$, 
%% ADDED %%
%
a growing solution $p^{+} _{{\rm II}}$ is ignored
since $\kappa r_{\rm out}/M \gg 1$.
Even if small value $p^{+} _{{\rm II}}(r_{\rm out})$ 
may be involved, the contribution 
exponentially decreases with the decrease of $r$.
The regular solution both at the horizon and at infinity
is given with a constant $c^{-} _{{\rm II}} $ by
\begin{equation}
p =   p^{\rm S}_{{\rm II}}  + c^{-} _{{\rm II}} p^{-} _{{\rm II}} .
\label{eqn.ptsol}
\end{equation}
The solution $h ~(\sim \chi^{-1} p)$ is obtained by the 
relation (\ref{eqn.typvv}), which may be unchanged within the lowest 
approximation.
Both $p^{\rm S}_{{\rm II}}$ and $ p^{-} _{{\rm II}} $ go to zero 
toward the horizon, $p^{\rm S}_{{\rm II}}, p^{-} _{{\rm II}} \propto 
\alpha^{1}  \to 0 $, whereas $h _{{\rm II}}$ diverges in general as
$h _{{\rm II}}  \propto  \alpha^{-1} $.
The regularity condition is
$\delta E_{\hat r} = - \delta \Phi_{, r} \sim \alpha^{0}$
and $\delta E_{\hat \theta} = -( \alpha r)^{-1}(\delta \Phi_{, \theta}-
\omega G_{,\theta} ) \sim \alpha^{-1}$
in ZAMO variables \citep{1986bhmp.book.....T}.
This means that the function $h _{{\rm II}}$ should be
finite at the horizon, and therefore
the constant $c^{-} _{{\rm II}} $ may be uniquely determined as
$c^{-} _{{\rm II}} =-(p^{\rm S}_{{\rm II}}/p^{-} _{{\rm II}})_{\alpha  \to 0}$.
That is, the divergence of 
$h _{{\rm II}} ( \propto  \alpha^{0})$ is suppressed by 
an appropriate of $ p_{{\rm II}} (\propto \alpha^{2} )$.
From now on, we consider this unique solution,
which we call the type II solution for brevity.

We next discuss the Znajek condition at the horizon.
The condition of eq.~(\ref{ZnajekCND}) can be written in terms of 
the first-order perturbed functions 
as
\begin{equation}
k p +  \chi ^{-1} h = -\frac{ a_{*}}{ 4 \chi} ,
\label{ZnajekCNhpa}
\end{equation}
where  $\delta \Phi = h \cos \theta$,
$\delta S =2 \lambda \delta F $
$= \lambda p \sin^2 \theta$, ${\bar G}=1-\cos\theta$,
$\delta  G=0$, and $\omega_{\rm H} =a_{*}/(4M)$ are used.
This condition is a relation between $p$ and $\chi ^{-1} h$, and 
can be easily incorporated in a type I WKB solution 
because their magnitudes are of order $ p^{\pm} _{{\rm I}}$ 
$ \sim \chi ^{-1} h^{\pm} _{{\rm I}}$. 
However, the ordering is opposite that of the type II solution, 
$ h^{\pm} _{{\rm II}}$ $ \sim  \chi ^{-1} p^{\pm} _{{\rm II}}$.
The leading-order WKB solutions are not satisfied with the 
condition (\ref{ZnajekCNhpa}), so that
we have to take into account the higher-order 
WKB approximation as 
$ p \propto \exp ( \chi W) \to  p_{(1)} 
  +\chi ^{-1} p_{(2)} +  \chi ^{-2} p_{(3)} + \cdots$, 
$ h \to  h_{(1)} +\chi ^{-1} h_{(2)} +  \chi ^{-2} h_{(3)} + \cdots$,
where the subscript ${(n)}$ denotes approximation order $n$. 
The condition (\ref{ZnajekCNhpa}) provides a relation between 
$ p_{(2)}$ and $ h_{(1)}$.
In the type II solution, we approximate
$ p_{(1)}=0$ at the horizon, but have to treat
a small correction $\chi ^{-1} p_{(2)} (\ll 1)$ in a more 
elaborate case.
%

%%3.4%%%%%%%%%%%%%%%%%%%%%%%%%%
\subsection{Results}
\begin{figure}
%\centering
\includegraphics[scale=0.75]{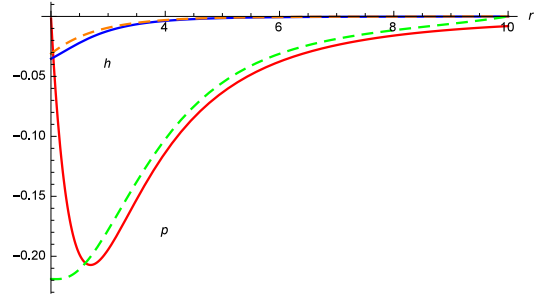}
\caption{ WKB solutions of $p$ and $h$ compared with numerical 
integration of eqs.~(\ref{eqn2x1})--(\ref{eqn2x2}).
The type II WKB solutions are plotted as solid curves, and the 
numerical results as dashed ones.
The parameters are chosen as $\lambda  =1$, $\chi  =25$ and
$a_{*} =1$.
}
\label{fig1}
\end{figure}

%%%%%%%%%%%%%%%%%%%%%%%%%%%%%%%
 We solve the differential eqs.~(\ref{eqn2x1}) and (\ref{eqn2x2})
with the perturbed Znajek condition of eq.~(\ref{ZnajekCNhpa})
at the horizon and $p = h=h_{, r} =0 $ 
at outer radius $r/M=10$.
%Fig1
The numerical solution is compared with the type II WKB solution $p$ 
given by eq.~(\ref{eqn.ptsol}) and the corresponding function $h$
by eq.~(\ref{eqn.typvv}).
Figure~\ref{fig1} shows good overall agreement except 
for the boundaries.
The agreement in $h$ is better. A minor difference around $r/M =10$ 
comes from the choice of outer boundary 
point; $r/M =10$ is chosen in the numerical integration,  
but it is infinity in the WKB solution.
There is another difference only in $p$ near the horizon.  
As discussed in the previous section, 
the function $p_{II}$ in leading-order WKB approximation goes to zero, 
but that of the numerical integration goes to a finite value, 
satisfying the Znajek condition.
Our WKB solution represents a piece of a numerical one,
which involves another type I WKB solution and higher corrections 
of $\chi^{-1} $.
Note that the Znajek condition (\ref{ZnajekCNhpa}) leads
to $p \approx 0$ at the horizon, if
the function $h $ is finite in large $\chi$ limit.

A set of differential eqs.~(\ref{eqn2x1}) and (\ref{eqn2x2})
is not so complicated as to prevent integration. 
However, numerical results cannot be obtained with 
high precision for large dimensionless parameters
because the characteristic lengths become very small.
For example, we found that typical upper limits are
$\chi < 50 $ and  $\lambda < 4 $.
Big or small number is involved in numerical integration like
e.g., 
$\exp( \pm \kappa s) \sim 10^4, 10^{-4}$ even for 
moderate case $\chi \sim 10 $ and  $\lambda \sim 1 $.  
Due to this limitation we use 
the type II leading-order WKB solution in place of numerical integration,
and explore the behavior in a much larger $\chi $ regime ($\chi \gg 1 $)
relevant to astrophysical situations. 
This also provides an advantage of easily analyzing parameter dependence.
%%%%%%%%%%%%%%%%%%%%%%%%%%%%%%%%%%%%%%%%%%%%%%%
%%%%% FIG1 %%%%%
%%%%%%%%%%%%%%%%%%%%%%%%%%%%%%%%%%%%%%%%%%%%%%%

\begin{figure}
%\centering
\includegraphics[scale=0.65]{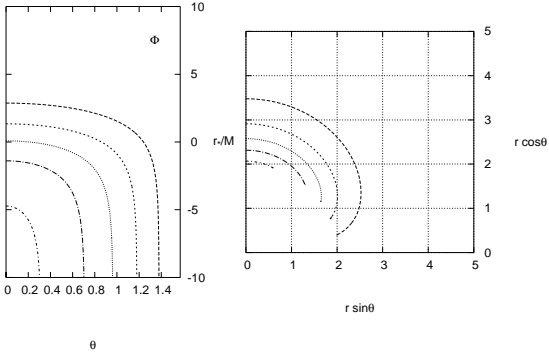}
\caption{Contour of electric potential $\delta \Phi (\le 0 )$
in the $\theta$-$r_*$ plane (left panel) and at the spherical
coordinate $( r, \theta )$ (right panel).
The minimum of $\delta\Phi $ is located on the polar axis
at the horizon $(\theta=0 , r=r_{\rm H} )$.
}
\label{fig2}
\end{figure}

%% FIG2 %%
Figure~\ref{fig2} shows the electric potential 
$\delta \Phi = h \cos \theta (\le 0)$ by the contours.
The minimum is located on the polar axis at the 
horizon ($\theta = 0$, $r/M \to 2$ or $r_*/M \to -\infty$), 
and $\delta \Phi $ increases with either the $\theta$ or the $r_{*}$
coordinate. The figure also shows that ${\vec \nabla} \delta \Phi$ 
is parallel to ${\vec \nabla} G$ near the horizon; explicitly,
$\delta \Phi_{,\theta} = -(h/M) G_{,\theta}$ in the $\theta$ direction.
Therefore, the function $h/M$ may be regarded 
as $\delta \Omega_{\rm F} \equiv -h/M$,
the angular velocity of the field line at the horizon. The velocity induced 
by first-order rotation is in the direction of black hole spin. 
A typical value of $\delta \Omega _{\rm F} $ for the model shown in Fig.~\ref{fig1} is $\sim 0.04a_{*}/M$.
By examining the behavior of $h$ in the WKB solution, 
we find that the value at the horizon is approximated as  
$h \approx -a_{*}/(2^5 \lambda \chi)^{1/2} $
$=-a_{*}/(2^{5/2} \kappa ) $.
As the number density (i.e., $\lambda$) increases, 
the deviation from the ideal MHD condition becomes small.
Consequently, we have $\delta \Phi \to 0$ in the 
large $\kappa$ limit.

\begin{figure}
\centering
 \includegraphics[scale=0.65]{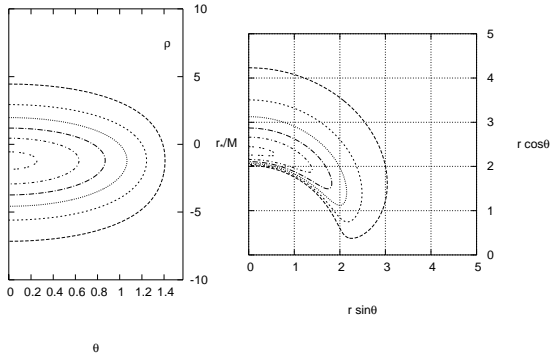}
\caption{Contour of 
$\delta n_+/ {\bar n}_0 (\propto \delta \rho_e)$
in the $\theta$-$r_*$ plane (left panel) and at the spherical
coordinate $( r, \theta )$ (right panel).
The minimum is located on the polar axis
$(\theta=0 , r/M \approx 2.2 )$. 
}
\label{fig3}
\end{figure}

%% FIG3 %%
  We now consider the density perturbation induced by 
black hole rotation. Figure~\ref{fig3} shows contours of 
$\delta n_{+}/ {\bar n}_0$
$=-\delta n_{-}/ {\bar n}_0$ ($\propto \delta \rho_{e}/ {\bar n}_0$)
in eq.~(\ref{eqn.denpert}). The minimum, negatively charged region 
for $a_{*}>0$, is located on the polar axis ($\theta = 0$).
Unlike $\delta \Phi$, the minimum is not on the horizon, 
but at $r/M \approx 2.2$. 
The rate of charge density to the background number density
goes to zero towards both radial directions
$r_* \to \pm \infty$ 
and becomes neutral at the horizon and infinity. 

%%%%%%%%%%%%%%%%%%%%%%%%%%%%%%%%%%%%%%%%%%%%%%%
%%%%% FIG2 %%%%%
%%%%%%%%%%%%%%%%%%%%%%%%%%%%%%%%%%%%%%%%%%%%%%%

%%%%%%%%%%%%%%%%%%%%%%%%%%%%%%%%%%%%%%%%%%%%%%%
%%%%% FIG3 %%%%%
%%%%%%%%%%%%%%%%%%%%%%%%%%%%%%%%%%%%%%%%%%%%%%%

\begin{figure}
%\centering
\includegraphics[scale=0.65]{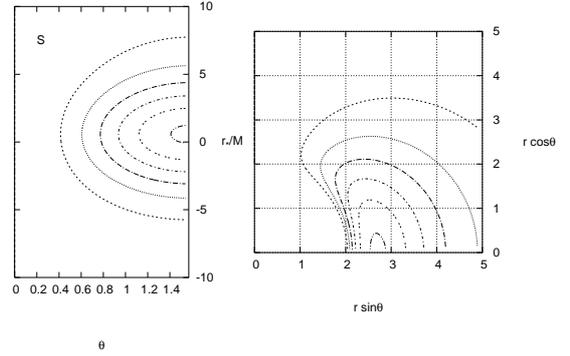}
\caption{ Contour of current function 
$\delta S/\lambda = 2\delta F (\le 0)$
in  the $\theta$-$r_*$ plane (left panel) and at the spherical
coordinate $( r, \theta )$ (right panel).
The minimum is located on the equator 
$(\theta=\pi /2,  r/M \approx 2.5)$.
}
\label{fig4}
\end{figure}

%% FIG4 %% 
 Figure~\ref{fig4} shows the stream function or current flow function   
$ 2 \delta F =\delta S/\lambda $
$  =  p \sin^2 \theta $ by contours.
Since $ \delta F <0 $,  
both fluids are dragged in the rotational direction of the black hole,
that is, $\delta v^{\pm} _{\hat\phi}  >0 $ (see eq.~(\ref{eqn.deltv3})). 
The effect is strong on the equator.
The contours also show the poloidal current flow $\delta S$, where
a half loop of current is formed around the minimum 
($\theta = \pi/2$, $r/M \approx 2.5$). 
In our treatment, we can never impose the condition 
$\delta S_{, r} =\lambda p_{,r}=0$ on the equator. 
This means that the poloidal current 
$j_{\theta} \propto -p_{,r}$
is allowed to flow into or out from the disk.
The function  $p_{,r}$ changes sign at $r/M \approx 2.5$.
The current emerges up ($j_{\theta} <0 $) from 
an outer point,  say,  $r_{2}/M >  2.5$, flows along a 
constant line of $\delta S$, and goes down to an inner point 
$r_{1}/M <  2.5$  (see eq.~(\ref{enqjcpol})).
However, there is no potential difference between $r_{1}$ and
$r_{2}$ on the equator. 
Rather, the origin is attributed to the current sheet on the equator, 
which produces the radial magnetic field in the vicinity of it.
The right panel in $(r, \theta )$ coordinates 
shows a strong concentration of contour lines near the horizon.
One might therefore assume a strong surface current on the horizon,
but,
as discussed in the previous section, the function $p$ in
a type II solution goes to zero as $ \sim \alpha ^{2}$
toward the horizon,
so there is no poloidal surface current on the horizon.
The left panel shows that the current vanishes
for $r_{*} \to -\infty$.
At the same time, the toroidal magnetic field 
$ \delta B_{\hat \phi} \sim \alpha $
also tends to vanish near the horizon.

%%%%%%%%%%%%%%%%%%%%%%%%%%%%%%%%%%%%%%%%%%%%%%%
%%%%% FIG4 %%%%%
%%%%%%%%%%%%%%%%%%%%%%%%%%%%%%%%%%%%%%%%%%%%%%%

%%
 We calculate the electromagnetic energy power through 
a sphere of radius $r$.
By integrating the perturbed expression of eq.~(\ref{EMpower}) 
with respect to $\theta$, we have
\begin{equation}
\delta P_{{\rm em}}(r) = \frac{2}{3} \lambda (B_0 M)^2 h p .
\label{EMpower2}
\end{equation}
The power $ \delta P_{{\rm em}}$ is positive for the type II solution,
because $ h p > 0$. In other words, outgoing flux is induced. 
In contrast, the functions for the type I solution
in which $ h p < 0$ represent incoming energy flux.

\begin{figure}
%\centering
\includegraphics[scale=0.8]{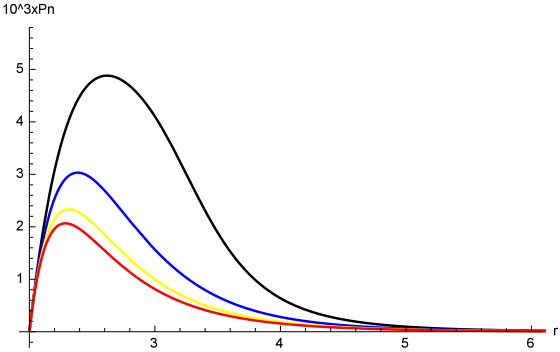}
\caption{ Outgoing electromagnetic power 
$ 10^{3} \delta P_{{\rm em}}/(a_{*}B_{0}M)^2$ as
a function of $r/M$.
Four models are shown with $k=0.025, 0.04, 0.1$ and $1$ 
from top to bottom.
}
\label{fig5}
\end{figure}

  Figure~\ref{fig5}
shows the outgoing power $ \delta P_{{\rm em}}$ 
as a function of $r$ for four models. 
The function increases with the decrease of $r$ and has
a maximum around $r/M =2.5$, 
where the longitudinal electric fields and toroidal magnetic fields
are significantly produced. However, the power declines 
toward the horizon and tends to zero owing to $p \to 0$.
This energy flow is therefore produced by the black hole
rotation, but is not related to the outgoing flux from the
horizon.
From numerical calculations, we found that 
$ \delta P_{{\rm em}} $ weakly depends on 
two parameters, ($\lambda $, $\chi$)
or ($\kappa^2 = \lambda \chi $, $k =\lambda /\chi$).

Figure ~\ref{fig5} shows that the peak slightly depends on $k$ 
and increases with $k^{-1}$, but does not depend on $\kappa$.  
The mathematical reason may be explained as follows:
The WKB expressions eqs.~(\ref{eqn.typvv}) and (\ref{eqn.wkb.ps}) 
yield $ p \sim k^{-1/2} $ and $ h \sim  \chi^{-1} k^{-1/2}  $,
so their product 
$ \delta P_{{\rm em}} $ 
$\sim \lambda \times k^{-1/2} \times \chi^{-1} k^{-1/2}$
$\sim  k^{0} $ in eq.~(\ref{EMpower2}), 
weakly depends on these parameters, despite $p$ and $h$ strongly 
depending on them in a different manner.
The electromagnetic fields perturbations
are approximated as 
$\delta B_{\phi} \sim \kappa  a_{*}B_{0} $
and 
$\delta \Phi \sim \kappa^{-1} a_{*} B_{0}$,
but the Poynting power, their product,
is almost independent of the parameter $\kappa$
in the microscopic level.

A typical value is of order $ \delta P_{{\rm em}}$
$\sim 5 \times 10^{-3} (a_{*} B_{0}M)^2$
$= $$8 \times 10^{-2} (a_{*} B_{\rm n}M)^2$,
where the magnetic field strength $B_{n} =B_{0}/4$ 
on the horizon is used.
Compare this number with the original estimate $ P_{{\rm BZ}}$ 
by Blandford--Znajek, 
calculated with the ideal MHD and force-free approximations
for a split monopole in the slow rotation limit.
The result depends on the undetermined angular velocity $\Omega_{\rm F}$
of the magnetic field lines. With the optimistic choice 
$\Omega_{\rm F} = \omega_{\rm H}/2 $, we have 
$P_{\rm BZ}= (a_{*} B_{\rm n}M)^2/6 
\sim 1.6 \times 10^{-1} (a_{*} B_{\rm n}M)^2$ in eq.~(\ref{positiveENG}).
We therefore have  $ \delta P_{{\rm em}} \sim 0.5 P_{\rm BZ}$.
These are of roughly the same order due to the ambiguity 
of $\Omega_{\rm F} $ involved in $P_{\rm BZ}$.
Finally, we calculate the conversion efficiency,
which is a ratio of the electromagnetic power 
to that of the background inflows given by eq.~(\ref{MATpower0}):
\begin{equation}
\frac{ \delta P_{{\rm em}}}{|P_{0}|}
\sim 2 \times 10^{-3} k^{-1} a_{*}^2.
\end{equation}
The efficiency increases with $k^{-1}$, and reaches
$\sim 0.1 a_{*}^2$ for $k \approx k_c=2.3\times 10^{-2}$.
This mechanism is efficient in strongly magnetized flow with less 
abundant matter, that is, smaller $k$.
Contrarily, the conversion becomes less active in high density 
cases, and may not work for $k>1$.
%

%%%%%%%%%%%%%%%%%%%%%%%%%%%%%%%%%%%%%%%%%%%%%%%
%%%%% FIG5 %%%%%
%%%%%%%%%%%%%%%%%%%%%%%%%%%%%%%%%%%%%%%%%%%%%%%

%(4)%%%%%%%%%%%%%%%%%%%%%%%%%%%%%%%%%%%%%%%%%%%
\section{Conclusion}
%%%%%%%%%%%%%%%%%%%%%%%%%%%%%%%%%%%%%%%%%%%%%%%
We have explored the electromagnetic
structure relevant to outgoing energy flux in 
a black hole magnetosphere.
Our treatment was based on a two-fluid description
in which the ideal MHD condition is no longer assumed.
By this formalism, we for the first time demonstrated 
how a longitudinal electric field is produced, even if
it vanishes at the outer boundary.
We first presented a general framework to construct a
stationary axially symmetric structure of the
electromagnetic fields and pair plasma flow.
The formalism is an extension of our previous
work \citep{2009MNRAS.398..271K} for pulsar magnetospheres
to a curved spacetime.

For a definite result, we limited discussion to a slowly rotating 
Kerr black hole and used the perturbation method with respect to 
the spin parameter.
We investigated how and where longitudinal electric fields and 
toroidal magnetic fields arise in the presence of black hole spin, 
although both fields are exactly zero in a spherical symmetric 
monopole model. To study the behavior 
at macroscopic lengths much larger than the plasma scales,
we used a WKB method.
By taking into account the first-order rotation, we found
a unique solution that describes zero fields at infinity and
finite longitudinal electric potential at the horizon.
This solution provides outgoing energy power,
with magnitude on the same order as that of the 
original work of \cite{1977MNRAS.179..433B}.

In comparison with that work,
we see a different physics involved 
in the plasma model. Split monopole magnetic fields as the 
lowest approximation and perturbations were used in both works. 
In both, the outgoing power originates from the black hole spin,
but the origin is not the same.
The black hole horizon, which is causally disconnected to 
the exteriors, plays a minor role;
the energy power generated in our model is 
damped toward the horizon.
The ergoregion is often considered as the origin of longitudinal 
electric fields and the outgoing power
\citep[e.g.,][]{2004MNRAS.350..427K,2014MNRAS.442.2855T}.
The relation to the existence of the ergoregion is unclear 
within this work because we can not treat it:The outer ergoregion 
radius coincides with the horizon, $2M$ in our lowest 
approximation of the spin.
As discussed in Section~2, the ideal MHD condition should be 
inevitably violated inside the ergoregion for a purely poloidal 
magnetic field, but this argument may be too strong.
The breakdown position is shifted outwardly 
in our pair model.
Finally, \citet{2000NCimB.115..795B}
discussed the importance of the pair creation surface
in determining the global MHD magnetosphere.
The radius in a slowly rotating split-monopole magnetosphere 
is $ \sim 2.52M$.
Inflows and outflows are modeled as the boundary of both sides.
In our model, the maxima of $|\delta \rho_{e}|$, 
$|\delta B_{\phi}|$, and the luminosity peak 
are located around $ r \sim$ 2.2--2.5$M$.
The pairs are assumed to exist outside the black hole,
and to fall into the horizon in the background mean flows.
We cannot treat outflows.
It is important to study the plasma behavior 
in the vicinity of the black hole horizon or ergosurface,
which leads to outgoing power by the dragging.
The crucial radius for the energy conversion is further 
pushed out by taking into account a realistic two-fluid  
Present result was derived under simplified conditions
such as steady state, split monopole, certain parameter ranges, 
boundary conditions and a leading-order WKB approximation.
It is true that further study is needed to confirm
outgoing Poynting power generation by carefully examining
each assumption.
%

%%%%%%%%%%%%%%%%%%%%%%%%%%%%%%%%%%%%%%%%%%%%%%%%%%
\section*{Acknowledgements}
%%%%%%%%%%%%%%%%%%%%%%%%%%%%%%%%%%%%%%%%%%%%%%%%%%
I am grateful to Kenji Toma and Katuma Kamitamari for useful discussions.
This work was supported in part by a Grant-in-Aid for Scientific Research
(No. 26400276) from the Japanese Ministry of Education, Culture, Sports,
Science and Technology.

%%%%%%%%%%%%%%%%%%%% REFERENCES %%%%%%%%%%%%%%%%%%
% \bibliographystyle{mnras}
% \bibliography{BZ2fld} 
%

%%%%%%%%%%%%%%%%%%%%%%%%%%%%%%%%%%%%%%%%%%%%%%%%%%

%%%%%%%%%%%%%%%%% APPENDICES %%%%%%%%%%%%%%%%%%%%%
 \appendix
%%%%%%%%%%%%%%%%%%%%%%%%%%%%%%%%%%%%%%%%%%%%%%%%%%
\section{Energy power}
%%%%%%%%%%%%%%%%%%%%%%%%%%%%%%%%%%%%%%%%%%%%%%%%%%

In this appendix, we consider stationary energy flow.
The energy momentum tensor $T^{\mu \nu}$ is a sum of electromagnetic 
($T_{{\rm em}}^{\mu \nu}$) and matter ($T_{(\pm)}^{\mu \nu}$) parts.
The outgoing energy per unit time through a constant radius $r$ 
is obtained by integrating 
the energy conservation equation
$(\sqrt{-g}T_t^{\mu})_{,\mu} /\sqrt{-g}=0$.
The power consists of a sum of the electromagnetic and matter parts: 
\begin{equation}
P_{{\rm em}}(r)+ P_{(+)}(r)+ P_{(-)}(r) =
- \int  (\sqrt{-g} T^r _t ) d\theta d\phi .
\label{totalpowerAp}
\end{equation}
This total power is independent of $r$,
but each part $P_{{\rm em}}$, $P_{(\pm)}$
in general depends on $r$ because of their interaction. 
The electromagnetic part $P_{{\rm em}}$,
that is, the outgoing Poynting flux, can be written as
\begin{equation}
P_{{\rm em}}(r)=- \int (\sqrt{-g} T_{{\rm em} ~t} ^r ) d\theta d\phi 
=-\frac{1}{2}\int (\Phi,_{\theta} S ) d\theta ,
\label{EMpowerAp}
\end{equation}
where the energy momentum tensor is 
expressed by electromagnetic fields measured by a ZAMO,
$T_{{\rm em} ~t} ^r=$
$-\Delta^{1/2}\rho^{-1}(\alpha E^{\hat{\theta}}
-\omega{\varpi}B^{\hat{r}})B^{\hat{\phi}}/{4\pi}$.

We next consider the plasma energy flow. Power in the $+r$ direction
is given for positively or negatively charged fluid components by
\begin{eqnarray}
%{equation}
P_{(\pm)}(r) &=& -\int  (\sqrt{-g} T_{(\pm) ~t} ^r ) d\theta d\phi 
\nonumber
\\
&= & 2\pi \int  (mK_{\pm} \mp e \Phi) F_{\pm ,\theta} d\theta, 
\label{MATpowerAp}
\end{eqnarray}
where  $T_{(\pm) ~t} ^r=$
$ [mn \gamma v^{r}(1+ \omega \varpi v_\phi/\alpha)]_{\pm} $
and Bernoulli integral $K_{\pm }$ (\ref{const.Bern})
are used.

The total power eq.~(\ref{totalpowerAp}),
a sum of eqs.~(\ref{EMpowerAp}) and (\ref{MATpowerAp}),
is expressed as 
\begin{equation}
P_{\rm em}+ P_{(+)}+ P_{(-)} =
-\frac{1}{2}\Big[ \Phi S \Big]_{\theta _{a}} ^{\theta _{b}} 
+ 2 \pi m \sum _{\pm} \int_{r=const}  K_{\pm} dF_{\pm } , 
\label{sumpower}
\end{equation}
where we integrated by parts and used the relation
$ S = 4\pi e(F_+ -F_-)$. 
The first term on the right-hand side in eq.~(\ref{sumpower}) denotes
the difference between the two boundary values with respect 
to $\theta$, and is zero in appropriate cases,
e.g., $S(0)= \Phi(\pi/2)=0$.
The second term on the right-hand side is constant with respect to $r$, 
unless the plasma flows go out through boundaries 
$\theta _{a}$ or $\theta _{b}$. 
We have thus confirmed that 
total energy flow is constant in a system consisting 
of an electromagnetic field and plasma.
We consider the right-hand side in eq.~(\ref{sumpower}) 
for the linearized perturbations considered in section 3.
The first term is also zero up to the second order, since 
the boundary value of $\delta \Phi S +\Phi \delta S 
+\delta \Phi \delta S $ becomes zero at $\theta=0, \pi/2$. 
We assumed that $K_{\pm}=1$. and $ \delta F_{+} = -\delta F_{-}$,
so that the sum in the second term becomes zero.
%
%%%%%%%%%%%%%%%%%%%%%%%%%%%%%%%%%%%%%%%%%%%%%%%%%%%%%%%%%%%%

\end{document}